\documentstyle[12pt]{article}

\newcommand{\be}{\begin{eqnarray}}
\newcommand{\ee}{\end{eqnarray}}

\newcommand\tag{\hbox to hsize}

\newcommand\del {\partial}

\input{tcilatex}

\begin{document}

\centerline{\Large\bf Yang-Mills Instantons in closed} \medskip %
\centerline{\Large\bf Robertson-Walker Space-Time$^\dagger$} \vskip3 cm %
\centerline{Hilmar Forkel} \vskip0.1cm \centerline{Physics Department} 
\centerline{State University of New York at
Stony Brook} \centerline{Stony Brook, New York, 11794.} \vskip1.5cm %
\centerline{NTG-89-48} \centerline{1989} \vskip2.5 cm \vskip1cm %
\centerline{\bf ABSTRACT} \vskip.5 cm

We construct a four-parameter class of self-dual instanton solutions of the
classical SU(2)-Yang-Mills equations in a closed Euclidean Robertson-Walker
space-time.

\vfil \noindent ${^\dagger}$Supported in part by the US Department of Energy
under Grant No. DE-FG02-88ER 40388 \eject

The discovery of the first Yang-Mills instantons \cite{bel75} and their
fundamental physical relevance \cite{cal76} initiated an intense research
activity devoted to their study. It led in particular to the construction of
multi-instanton solutions in continuous families with an increasing number
of degrees of freedom. Finally, with the advent of the ADHM method \cite%
{ati78} an explicit construction principle for the most general self-dual
solutions has been given, and as a consequence \cite{bo79} all stable
finite-action solutions of the Euclidean Yang-Mills equations in flat
(compactified) space-time are known today.

For instantons in curved space-time, however, the situation is quite
different. No general classification exists and not many explicit solutions
are known. Soon after the first flat space-time solution a straightforward
generalization to conformally flat de-Sitter space-times was found using the
conformal invariance of the self-duality equation \cite{egu76},\cite{dea79}.
Shortly afterwards a one-instanton solution in the Schwartzschild geometry,
suitably continued to Euclidean signature, was constructed \cite{cha177}. It
showed in particular that even very weak gravitational fields can have a
nontrivial influence on the general topological structure of instantons.
Further explicit Yang-Mills instanton solutions are known in $S^4$
space-times \cite{dol81} and in the Taub-NUT metric \cite{pop78}.

In this paper we are looking for new instantons in a curved Euclidean
space-time with topology $S^3 \otimes R$. We first discuss the general
implications of this topology for structure and classification of the
instantons. We then construct a four-parameter class of explicit solutions
of the self-duality equation for the Yang-Mills field strength and summarize
with a brief discussion.

The topological structure of nontrivial classical configurations of an SU(2)
gauge field $A$ is determined by the physical requirement of a finite
Euclidean Yang-Mills action 
\begin{eqnarray}
S = - \frac{1}{{\rm g}^2} \int_M tr[F \wedge *F],  \label{action}
\end{eqnarray}
where $M$ is a four dimensional Euclidean space-time and 
\begin{eqnarray}
F = dA + A \wedge A = \frac12 (\del_{\mu} A_{\nu} - \del_{\nu} A_{\mu} +
[A_{\mu},A_{\nu}] )\, dx^{\mu} \wedge dx^{\nu}
\end{eqnarray}
is the field strength of the (SU(2)-Lie algebra valued) gauge field, $*F$
its dual and g is the Yang-Mills coupling constant. (The trace is over the
internal indices.)

We choose the topological structure of space-time to be $S^3 \otimes R$ (the
three compact dimensions are spacial), but do not yet fix the detailed form
of the metric aside from its Euclidean signature and the topology. The
spacial hypersphere at time $t$ will be denoted $S^3_t$.

The finite-action condition forces the field strength $F$ to vanish
sufficiently fast in the asymptotic regions $S^3_t$ with $t \rightarrow \pm
\infty$ of space-time, in which $A$ has consequently to become pure gauge: 
\begin{eqnarray}
A \rightarrow \quad U_{\pm \infty}^{-1} dU_{\pm \infty} \hskip 1.5 cm {\rm %
for}\,\,\,\,\, t \rightarrow \pm \infty.  \label{bc}
\end{eqnarray}
Every continuous gauge-field configuration with finite action defines in
this way two mappings $U_{-\infty}, U_{+\infty}$ from the asymptotic regions 
$S^3_{t= -\infty}$ and $S^3_{t= +\infty}$, respectively, to SU(2) $\cong S^3$%
.

Both these mappings fall into homotopy classes, which form the third
homotopy group of $S^3$, $\pi_3(S^3) = Z$, and are therefore characterized
by integer degrees (winding numbers) $q_{-\infty}$ and $q_{+\infty}$. So
each gauge-field configuration specifies values for both these integers,
which are unchanged by any continuous deformation of the field.

Topologically nontrivial gauge transformations (which can be nonsingular for
the considered space-time topology), however, may change ($q_{-\infty}$, $%
q_{+\infty}$). Every such transformation maps $S_t^3$ to $SU(2)$ for all $t$
and by continuity always with the same homotopy degree . A gauge
transformation with nonzero degree changes the asymptotic form of $A$, eq. (%
\ref{bc}), and in particular adds its degree to the degrees of $U_{\pm
\infty}$. This leads to the conclusion that only the difference of these
degrees, $Q = q_{+\infty}$ - $q_{-\infty}$, characterizes a field
configuration in a gauge-independent way. Indeed, $-Q$ can be identified
with the integral of the second Chern class of $F$ over $M$, 
\begin{eqnarray}
Q = - \frac{1}{8 {\pi}^2} \int_M tr[F \wedge F],  \label{chern}
\end{eqnarray}
which is exactly the quantity used for the topological classification of
gauge fields in flat space-time. It is gauge invariant and well defined for
the asymtotically decaying $A$'s under consideration. In order to show its
equivalence with the first definition of $Q$ explicitly, we use\footnote{%
The integrand in (\ref{chern}) being an exact form (total divergence) makes $%
Q$ a topological invariant on a compact manifold without boundary, the
second Chern number.} 
\begin{eqnarray}
tr[F \wedge F] = d tr(F \wedge A - \frac13 A \wedge A \wedge A)
\end{eqnarray}
together with Stoke's theorem (assuming $A$ to be nonsingular on $M$) in
order to rewrite $Q$ as a surface integral over the boundary of $S^3 \otimes
[-\tau,\tau]$ for $\tau \rightarrow \infty$, {\it i.e.} in a very large
compact time interval of $M$, 
\begin{eqnarray}
Q = - \frac{1}{8 {\pi}^2} \int_{\del M} tr(F \wedge A - \frac13 A \wedge A
\wedge A)
\end{eqnarray}
with 
\begin{eqnarray}
\del M = S^3_{+\tau} + S^3_{-\tau} \qquad {\rm for}\, \tau \rightarrow
\infty.
\end{eqnarray}
Using the asymptotic form of $A$ on $\del M$, (\ref{bc}), we arrive as
anticipated at 
\begin{eqnarray}
Q = q_{+\infty} - q_{-\infty},  \label{inr}
\end{eqnarray}
where 
\begin{eqnarray}
q_{\pm \infty} = \lim_{\tau \rightarrow \pm \infty} \frac{1}{24\pi^2}
\int_{S^3_{\tau}} tr[U_{\pm \infty}^{-1} dU_{\pm \infty} \wedge U_{\pm
\infty}^{-1} dU_{\pm \infty} \wedge U_{\pm \infty}^{-1} dU_{\pm \infty} ]
\label{wind}
\end{eqnarray}
is the well-known expression for the homotopic degree of the mappings $%
U_{\pm \infty}$, respectively. The gauge invariance of $Q$ can now be
checked explicitly by using the behaviour of (\ref{wind}) under gauge
transformations.

We are now going to construct explicit instanton solutions of the Yang-Mills
equations in the $Q=1$ sector, which provide an example for the general
asymptotic structure of the gauge field and its topological classification.
In particular we are interested in solutions with self-dual field strength 
\begin{eqnarray}
F = *F,  \label{selfd}
\end{eqnarray}
which correspond to absolute minima of the action\footnote{%
By using Schwarz's inequality for the inner product of Lie-algebra valued
2-forms on $M$ one establishes as in flat space the lower bound $S \geq
8\pi^2 |Q|/ {\rm g}^2 $ for the action in the topological sector $Q$, which
is realized by self-dual fields.}. Note, however, that not all stable
finite-action solutions of the Yang-Mills equations have to be necessarily
self-dual in curved space-time.

To find solutions of (\ref{selfd}) with $Q$ = 1, we start with the ansatz 
\begin{eqnarray}
A = f(t) \, U_1^{-1} d U_1, \qquad U_1 = \cos \psi + i\tau_a{\hat r}%
^a(\theta,\phi) \sin \psi  \label{ansatz}
\end{eqnarray}
($\theta,\phi$ and $\psi$ are polar coordinates on the spacial $S^3$ of $M$, 
${\hat r}^a$ are the cartesian components of the unit vector in $R^3$ and $%
\tau_a$ are the Pauli matrices), which preserves the rotational (equivalent
to constant gauge-) symmetry of the asymptotic form (\ref{bc}). Note that
this ansatz gives $A$ in temporal (Weyl-) gauge, $A_0 = 0$. $U_1$ is
(topologically) the identity map from the spacial $S^3$ to SU(2). Inserting (%
\ref{ansatz}) into (\ref{selfd}), the self-duality condition becomes 
\begin{eqnarray}
\dot{f} L_i = \frac12 f(f-1) \epsilon_{ijk} [L^j,L^k].  \label{selfd2}
\end{eqnarray}
The latin indices ($i=1,2,3$) refer to the spacial coordinates on $M$, the
dot indicates the time derivative and $L_i = U_1^{-1} \del_i U_1 =
\theta^a_{\,\,\,i}\, i\tau_a$ are the components of the left-invariant
Cartan-Maurer form, pulled back by the identity map $U_1$. ($\epsilon_{ijk}$
are the components of the Levi-Civita (volume) form on the spacial
hypersphere.)

If (\ref{selfd2}) is to hold for all times, the time independence of the $%
L_i $'s forces 
\begin{eqnarray}
\alpha = f(f-1)/\dot{f}  \label{alpha}
\end{eqnarray}
to be constant. The nonsingular solution of this differential equation, 
\begin{eqnarray}
f(t) = \frac{1}{1 + \exp{(t-t_0)/\alpha} },  \label{f}
\end{eqnarray}
determines the time dependence of the instanton. The integration constant $%
t_0$ reflects the translational invariance of the action (\ref{action}) in
time direction.

After decomposing the spacial part of the metric 
\begin{eqnarray}
ds^2 = g_{\mu \nu} dx^{\mu} dx^{\nu} = dt^2 + d\sigma^2  \label{metric}
\end{eqnarray}
on $M$ into an orthonormal dreibein field, 
\begin{eqnarray}
d\sigma^2 = \gamma_{ij} dx^i dx^j, \qquad \gamma_{ij} = \delta_{ab}
e^a_{\,\,\,i} e^b_{\,\,\,j},  \label{s3met}
\end{eqnarray}
(latin indices from the beginning of the alphabet label the axis of the
orthonormal frame), eqn. (\ref{selfd2}) can be rewritten as 
\begin{eqnarray}
[L_a,L_b] = \alpha^{-1} \epsilon_{ab}^{\,\,\,\,\,\,\,c} L_c.  \label{R}
\end{eqnarray}
The matrices $\,\alpha L_a(x) = O_a^{\,\,\,b}(x) \frac{\tau_b}{2i}$ must
therefore form locally a representation of the basis of the SU(2) Lie
algebra, which implies the orthogonality of $O_a^{\,\,\,b}(x) = -2 \alpha
e_a^{\,\,\,i} \theta^b_{\,\,\,i}$. Redefining the dreibein by a rotation
with $O^{-1}$ (which doesn't affect the decomposition of the metric (\ref%
{s3met})), its inverse becomes proportional to the pulled-back Cartan-Maurer
form: 
\begin{eqnarray}
e^a_{\,\,\,i} = -2 \alpha \theta^a_{\,\,\,i}.  \label{es}
\end{eqnarray}
Now the $\theta$'s are the pull-back of an orthonormal frame for the natural
left-invariant metric on SU(2) (the three-dimensional geometrical unit
sphere), which are still proportional to an orthonormal frame on the spatial
hypersphere, as (\ref{es}) shows. This implies that the spacial metric on $M$
is that of a {\it geometrical} three-sphere, too. Indeed, 
\begin{eqnarray}
d\sigma^2 = 4 \alpha^2 \delta_{ab} \theta^a_{\,\,\,i} \theta^b_{\,\,\,j}
\,dx^i dx^j = 4 \alpha^2 [d\psi^2 + \sin^2\psi (d\theta^2 + \sin^2\theta
d\phi^2)]
\end{eqnarray}
is the metric of a three-dimensional sphere with radius ${\cal R} =
2|\alpha| $ in polar coordinates. The complete space-time metric is
therefore the Euclidean analog of a (homogenous and isotropic)
Robertson-Walker line element with a constant spacial radius.

The above derivation shows that solutions of the self-duality equation of
the form (\ref{ansatz}) in space-times with topological structure $S^3
\otimes R$ exist only if space is a static geometrical hypersphere. The
metric (\ref{metric}) is therefore uniquely determined to be the Euclidean
version of the Einstein universe \cite{ein17}.

The preceding discussion determined only the modulus of $\alpha$, $%
\,|\alpha| = {\cal R}/2$. Inspection of (\ref{alpha}) and (\ref{es}) shows
that the two possible signs interchange the asymptotic values of $f$ as well
as the orientation of the $e^a_{\,\,\,i}$ and lead to two different
instanton solutions\footnote{%
This is a consequence of the PT invariance of the Yang-Mills action and its
spontaneous breaking by the instanton.}. They can, however, be transformed
into each other by $U_1$, eq. (\ref{ansatz}), taken as a (topologically
nontrivial) gauge transformation. We will choose the solutions with negative 
$\alpha$ as the representatives of their gauge-equivalence class and denote
them as $A^{+}$.

Eq. (\ref{selfd2}) shows that anti-self-dual solutions $A^{-}$, the
anti-instantons with $Q=-1$, can be obtained from the instanton solutions by
changing $\alpha$ to $-\alpha$ in the definition of $f$, eq. (\ref{f}).

Implementing these remarks, the explicit form of the solutions in a closed,
static Robertson-Walker space-time of radius ${\cal R}$ finally becomes 
\begin{eqnarray}
A^{\pm} = \frac{1}{1 + \exp{\mp 2(t-t_0)/{\cal R}}) } U_1^{-1} d U_1
\end{eqnarray}
with $U_1$ as in (\ref{ansatz}).

They form a continuous four-dimensional family, parametrized by their center
in time direction, $t_0$, and three parameters corresponding to a global
SU(2) right transformation, which merely changes the (arbitrary) relative
position of the poles on $S^3_{space}$ and $S^3_{SU(2)}$.

The field strength of these (anti-) instantons, expressed in the orthonormal
frame, is 
\begin{eqnarray}
F^{\pm} = \dot{f} \frac{\tau_a}{2i \alpha} (e^0 \wedge e^a \pm \frac12
\epsilon^a_{\,\,\,bc} \, e^b \wedge e^c),
\end{eqnarray}
where $e^a = e^a_{\,\,\,i}dx^i$ and $e^0 = dt$. Note that the extension of
the instanton in time $|\alpha| \sim {\cal R}$ as well as in space is
gouverned by ${\cal R}$. In contrast to the flat space case, the metric
introduces a dimensionful parameter which fixes the scale of the solutions.

The second Chern class of the (anti-) instantons, 
\begin{eqnarray}
c_2^{\pm} = \frac{1}{8 \pi^2} tr[F^{\pm} \wedge F^{\pm}] = \mp \frac{\dot{f}%
^2}{16 \pi^2 \alpha^2} \epsilon_{abc} e^0 \wedge e^a \wedge e^b \wedge e^c,
\end{eqnarray}
can be used to check their instanton number explicitly: 
\begin{eqnarray}
Q^{\pm} = - \int_M c_2^{\pm} = \pm 3 {\cal R} \int_{-\infty}^{\infty} \dot{f}%
^2 \,dt = \pm 1.
\end{eqnarray}

To summarize, we have presented a four-parameter family of self-dual
instanton solutions of the SU(2)-Yang-Mills equation in a closed, static
Ro\-bert\-son-Wal\-ker space-time. Because of the well-known fact that the
Eu\-cli\-dean energy-momentum tensor of self-dual fields vanishes, they do
not interact (at the classical level) with gravity. The metric (\ref{metric}%
) together with the instantons may consequently be interpreted as a
self-consistent classical solution of the Einstein-Yang-Mills equations with
space-time in form of a matter-dominated Euclidean Einstein universe.

Furthermore, the obtained solutions provide a new example for the influence
of an even very weak gravitational space-time curvature on the qualitative
structure and scale of Yang-Mills instantons. In the light of the recent
investigations of Skyrmions on three-dimensional hyperspheres \cite{man87},
they might also, as in flat space, be useful in providing new Skyrmion
configurations and their generalization to finite temperature \cite{now89}
via the Atiyah-Manton construction \cite{ati89}.

We would like to thank Andy Jackson for raising the question of instantons
in a hyperspherical space and the scientific council of NATO for a research
fellowship.

\newpage


\newpage

\begin{thebibliography}{99}
\bibitem{bel75} A.A. Belavin, A.M. Polyakov, A.S. Schwarz and Yu.S. Tyupkin,
Phys. Lett. {\bf 59B} (1975) 85

\bibitem{cal76} C.G. Callan, R. Dashen and D.J. Gross, Phys. Lett. {\bf 63B}
(1976) 334; G. t'Hooft, Phys. Rev. Lett. {\bf 37} (1976) 8, Phys. Rev. {\bf %
D14} (1977) 3432

\bibitem{ati78} M.F. Atiyah, N.J. Hitchin, V.G. Drinfeld and Yu.I. Manin,
Phys. Lett. {\bf 65A} (1978) 185

\bibitem{bo79} J.P. Bourguignon, H.B. Lawson and J. Simons, Proc. Nat. Acad.
Sci. {\bf 76} (1979) 1550

\bibitem{egu76} T. Eguchi and P.G.O. Freund, Phys. Rev. Lett. {\bf 37}
(1976) 1251

\bibitem{dea79} V. de Alfaro, S. Fubini and G. Furlan, Nuovo Cimento {\bf 50A%
} (1979) 523

\bibitem{cha177} J.M. Charap and M.J. Duff, Phys. Lett. {\bf 69B} (1977)
445; {\bf 71B} (1977) 219

\bibitem{dol81} B.P. Dolan, J. Phys. {\bf A14} (1981) 1205

\bibitem{pop78} C.N. Pope and A.L. Yuille, Phys. Lett. {\bf 78B} (1978) 424

\bibitem{ein17} A. Einstein, {\it Sitzungsberichte der Preu{\ss}ischen
Akademie der Wissenschaften} 1917, p. 142

\bibitem{man87} N.S. Manton, Commun. Math. Phys. {\bf 111} (1987) 469; A.D.
Jackson, NORDITA preprint 87-38 N, 1987

\bibitem{now89} M. Nowak and I. Zahed, Stony Brook preprint, 1989

\bibitem{ati89} M.F. Atiyah and N.S. Manton, Cambridge preprint DAMTP-89-07
\end{thebibliography}
\end{document}